\documentclass[a4paper,superscriptaddress,aps,prb,twocolumn,floatfix,citeautoscript,longbibliography]{revtex4-2}
\usepackage{bbold}

\usepackage{dsfont}
\usepackage[utf8]{inputenc}
\usepackage{amsmath}
\usepackage{amssymb}
\usepackage{graphicx}
\usepackage{cleveref}
\usepackage{xcolor}
\usepackage{cancel}
\usepackage{booktabs}
\usepackage{ulem}

\usepackage{feynmp-auto}
\usepackage{lipsum}
\usepackage{lmodern}
\usepackage{tikz}
\usetikzlibrary{decorations.pathmorphing}

\tikzset{snake it/.style={decorate, decoration={snake, segment length=1.5mm,amplitude=0.5mm}}}

\newcommand{\lcpq}{Univ Toulouse, CNRS, Laboratoire de Chimie et Physique Quantiques, Toulouse, France}
\newcommand{\lpt}{Univ Toulouse, CNRS, Laboratoire de Physique Th\'eorique, Toulouse, France}
\newcommand{\etsf}{ European Theoretical Spectroscopy Facility, Toulouse }

\begin{document}
\title{The multichannel Dyson equation for double ionisation spectroscopies}
\author{Pierre Selli\' e}
\email{pierre.sellie@irsamc.ups-tlse.fr}
\affiliation{\lcpq}
\affiliation{\lpt}
\affiliation{\etsf}
\author{J. Arjan Berger}
\email{arjan.berger@irsamc.ups-tlse.fr}
\affiliation{\lcpq}
\affiliation{\etsf}
\author{Pina Romaniello}
\email{pina.romaniello@irsamc.ups-tlse.fr}
\affiliation{\lpt}
\affiliation{\etsf}

\begin{abstract}
Several photoemission spectroscopies and, in particular, Auger spectroscopy, involve double-ionization processes.
For the numerical simulation of these spectroscopies it is convenient to use the particle-particle channel of the two-body Green's functions
since its poles correspond to excitation energies in which the final state has two more particles (holes or electrons) than the initial state.
In standard approaches it is approximated within the random phase approximation.
As a consequence only the quasiparticles of the photoemission spectrum are captured but none of the satellites features.
In this work, we go beyond this approximation by employing the multichannel Dyson equation. 
By coupling the particle-particle two-body Green's function to the 3-hole-1-electron and 3-electron-1-hole channels of the four-body Green's function, the multichannel Dyson equation incorporates correlations beyond the RPA in a straightforward way.
We are thus able to describe both quasiparticles and satellites in the photoemission spectra.
\end{abstract}

\maketitle
\section{Introduction}
Ionization spectroscopy techniques that lead to the formation of doubly charged ions, such as Auger spectroscopy or direct double photoionization, play a central role in modern chemical analysis due to their sensitivity to electron-electron interactions, chemical environment, and local atomic structure. However, double ionization is a highly complex process, and the resulting spectra contain many closely spaced states, making them difficult to interpret without solid theoretical guidance. In this context, employing theoretical formalisms based on so-called Green's functions (GF) is highly useful, as these methods directly target spectroscopic observables. While the one-body Green's function is associated with photoemission and inverse photoemission spectroscopies, which probe the removal or addition of a single electron, the two-particle Green's function provides the appropriate framework for describing double-ionization processes, in which two electrons are removed. However, unlike the electron-hole (eh) channel, which has been widely used and developed by the solid-state community and more recently adopted by the quantum chemistry community, the pp channel has received relatively less attention. Most existing applications have been formulated within the random phase approximation (RPA) or its screened extensions \cite{PhysRevB.80.155201,PhysRevB_Stefanucci,Nat_Commun,Deilmann_2016,Torche_2019,Torche_2021,Steinhoff_PRX}. Notably, extensions beyond RPA have been developed within the nonequilibrium Green's function formalism based on the Kadanoff-Baym equations \cite{Bornath_1999}. The pp channel has also been explored within the algebraic diagrammatic construction (ADC) framework. \cite{Schirmer_1984,Tarantelli_2006} In particular, ADC(2) has proven to be an essential advancement beyond RPA and has been successfully applied to various molecules.\cite{Schirmer_1984,Tarantelli_1994,Villani_2004,Feifel_2005,Feyer_2005} Besides the study of double ionization potentials, recent studies have used the particle-particle RPA (pp-RPA) as a promising alternative for accessing neutral excitations. Indeed, the excitation energies of a $N$-electron system can be computed as the differences between the two-electron addition (DEA) energies of the (N$-$2)-electron system or, alternatively, from the differences between the two-electron removal energies of the ($N + 2$)-electron system.\cite{Yang_2013,Yang_2014,Yang_2015,Yang_2016,Yang_2017,Li_2024_1,Li_2024_2,Bannwarth_2020} It has been shown that pp-RPA can overcome some of the challenges faced by TDDFT and eh-RPA, such as the description of double excitations, Rydberg excitations, charge-transfer (CT) excitations, and conical intersections. 
However, despite its advantages, pp-RPA does not always provide sufficient accuracy, which has prompted the development of methods that extend beyond the pp-RPA framework. In this context, Marie \textit{et al.}\cite{Marie_2025} developed approximations beyond RPA for the pp channel by considering pairing fields and anomalous Green's functions. This results in a pp Bethe-Salpeter equation (BSE) and approximations to its kernel that are analogous to the eh counterpart.

In this paper, we take a different route.
We derive approximations that go beyond the RPA using the multichannel Dyson equation (MCDE) for double ionization and double addition.
Originally developed as an alternative to Hedin's $GW$ approximation to describe direct and inverse photoemission spectroscopy 
\cite{Riv22,riva_prl,riva_prb,paggi_2025}, the MCDE was later successfully extended to neutral excitations \cite{riva_prb_25} and now represents a general theoretical framework.
The main idea of the MCDE is to couple two (or more) independent-particle Green's functions that correspond to the same final state.
For example, by coupling the 1-body Green's function to the 2-hole-1-electron and 2-electron-1-hole channels of the 3-body Green's function one can accurately describe both quasiparticles and satellites in photoemission spectra.
In this work, we exploit the systematic framework of the MCDE to construct approximations beyond the RPA for double ionization and double addition.

The article is organized as follows. In Sec.~\ref{Theory} we show the link between the 3-hole-1-electron and 3-electron-1-hole channels of the 4-body Green's function and the particle-particle channel of the two-body Green's function. We then derive the multichannel Dyson equation that couples these channels through a 4-body self-energy, for which we derive a simple approximation. We finally map the multichannel Dyson equation onto an eigenvalue problem which can be easily solved using standard numerical techniques. 
We finally draw our conclusions and perspectives in Sec.\ref{Conclusions}.

%

%
\section{Multichannel Dyson equation for two-electron removal and addition\label{Theory}}
In this section we will derive the multichannel Dyson equation corresponding the removal and addition of two electrons.
It is based on the coupling of the particle-particle channel of the 2-body Green's function and the 3-hole-1-electron and 3-electron-1-hole channels of the 4-body Green's function.
\subsection{The particle-particle Green's function}
The spectral representation of the particle-particle (pp) channel of the two-body Green's is given by\cite{Stri88}
\begin{align}
& G_2^{2p}(x_1,x_2,x_{1'},x_{2'};\omega) =-
      i\lim_{\eta \to 0^+} \sum_n\nonumber\\
&\Bigg[\frac{X_n(x_{1},x_2)\Tilde{X}_n(x_{1'},x_{2'})}{\omega - (E_n^{N+2} - E_0^N) + i\eta} 
-\frac{\Tilde{Z}_n(x_{1},x_2)Z_n(x_{1'},x_{2'})}{\omega + (E_n^{N-2} - E_0^N) - i\eta}\Bigg]
        \label{Eqn:G2pextactdef}
\end{align}
with
\begin{align}\label{Eqn:G2_amplitudes}
          X_n(x_{1},x_2)&= \langle\Psi_0 ^N|\hat{\psi}(x_{1})\hat{\psi}(x_2)|\Psi_n ^{N+2}\rangle , \nonumber \\  
          \Tilde{X}_n(x_{1},x_{2}) &=\langle\Psi_n ^{N+2}|\hat{\psi}^\dag(x_2)\hat{\psi}^\dag(x_1)|\Psi_0 ^N\rangle \nonumber\\
          Z_n(x_{1},x_2)&= \langle\Psi_0 ^N|\hat{\psi}^\dag(x_2)\hat{\psi}^\dag(x_1)|\Psi_n ^{N-2}\rangle , \nonumber \\  
          \Tilde{Z}_n(x_{1},x_{2}) &=\langle\Psi_n ^{N-2}|\hat{\psi}(x_{1})\hat{\psi}(x_2)|\Psi_0 ^N\rangle
\end{align}
and $E_n^{N\pm 2}$ the eigenvalues of the $(N\pm 2)$-electron system.

The removal or addition of two electrons perturbs the system and can lead to the creation of electron-hole pairs. These excitations correspond to distinct features in double-ionization and double-addition spectra called satellites. \cite{Ferte_2017,Feyer_2005} They are included in $G_2^{2p}(\omega)$ but completely absent in an independent-particle approximation of the pp Green's function $G_2^{0,2p}(\omega)$, such as the Hartree-Fock pp Green's function.
Therefore, when solving the single-channel Dyson equation for $G_2^{2p}(\omega)$, given by
\begin{equation}\label{Dyson2:eq}
  G_2^{2p}(\omega)=G_2^{0,2p}(\omega)+ G_2^{0,2p}(\omega) \Sigma_2^{pp}(\omega)  G_2^{2p}(\omega),
\end{equation}
the satellites have to be included through a frequency-dependent self-energy $\Sigma_2^{pp}(\omega)$ 
for which it is nontrivial to find good approximations.

Instead, as we will show in the following, even in the independent-particle approximation, the 3-hole-1-electron and 3-electron-1-hole channels of the 4-body Green's function already contain information about satellites.
As a consequence this hugely simplifies finding good approximations of the corresponding self-energy.


%
\subsection{The 4-body Green's function}
The 4-GF is defined by 
\begin{align}\label{G4def:eq}
    & G_4(1,2,3,4,1',2',3',4') = 
    \nonumber \\ & \langle\Psi_0 ^N|\hat{T}[\hat{\psi}(1)\hat{\psi}(2)\hat{\psi}(3)\hat{\psi}(4)\hat{\psi}^\dag(4')\hat{\psi}^\dag(3')\hat{\psi}^\dag(2')\hat{\psi}^\dag(1')]|\Psi_0 ^N\rangle,
\end{align}
with $1=(x_1,t_1)$, where $x_1=(\mathbf{r}_1,\sigma_1)$, $|\Psi_0^N\rangle$ the ground-state many-body wavefunction of an $N$-electron system, and $\hat{T}$ the time ordering operator. Different choices of the time ordering yield different orders of the field operators and, therefore, different physical information. The 4-GF describes the propagation of four particles (electrons or holes) and it can be split into five components: $G_4^{4e}$, $G_4^{3e1h}$, $G_4^{2e2h}$, $G_4^{1e3h}$ and $G_4^{4h}$.
In this work we focus on processes with a final state that has two particles (electrons or holes) more than the initial ground state. For this reason, in the following, we focus ot the $3e1h$ and $3h1e$ components of the 4-GF.
We will show that by coupling it to the $2p$ channel of the 2-GF we can treat quasiparticles and satellites on equal footing.
In order to isolate the $3e1h$ and $3h1e$ channels of the 4-GF we choose the following time differences 
\begin{align}\label{timesG4:eq}
   && \tau_{12}=0^+, && \tau_{23}=0^+, && \tau_{31'}=0^+,&& \tau_{44'}=0^+,\nonumber\\  && \tau_{4'3'}=0^+, && \tau_{3'2'}=0^+, && \tau=t_1-t_3,
\end{align}
where $\tau_{ij}=t_i-t_j$.
The time difference $\tau$ corresponds to the simultaneous propagation of the four particles in the system. The other time differences vanish, which corresponds to the simultaneous creation (and destruction) of 
the four particles.
We thus obtain the following expression for $G_4^{3h1e/3e1h}$,


\begin{align}
 &G_4^{3h1e/3e1h}(x_1,x_2,x_3,x_4,x_{1'},x_{2'},x_{3'},x_{4'};\tau)    = \nonumber\\
 &\langle\Psi_0 ^N|T[(\hat{\psi}(x_1)\hat{\psi}(x_2)\hat{\psi}(x_{3})\hat{\psi}^{\dagger}(x_{1'}))_{t_1}
    \nonumber \\ & \times
    (\hat{\psi}(x_4)\hat{\psi}^\dag(x_{4'})\hat{\psi}^\dag(x_{3'})\hat{\psi}^\dag(x_{2'}))_{t_3}]|\Psi_0 ^N\rangle,
    \label{Eqn:3h1e-3e1hG3}
\end{align}
where the subscripts $t_1$ and $t_3$ on the right-hand side of Eq.~\eqref{Eqn:3h1e-3e1hG3} indicate that all the operators in the brackets act at time $t_1$ and $t_3$, respectively.
By introducing the closure relation in Fock space $\sum_M \sum_k | \Psi_k^{M}\rangle \langle\Psi_k^{M} | =\mathbb {1}$ in Eq.~\eqref{Eqn:3h1e-3e1hG3}, 
where $|\Psi_k^{M}\rangle$ indicates the $k$-th eigenstate of the $M$-electron system, and Fourier transforming with respect to $\tau$ leads to the following spectral representation
\begin{align}\label{G4spectral:eq}
        &G_4^{3h1e/3e1h}(x_1,x_2,x_3,x_4,x_{1'},x_{2'},x_{3'},x_{4'};\omega) =
        \nonumber \\ & i\lim_{\eta \to 0^+} \sum_n \big[ \frac{X_n(x_{1},x_2,x_{1'},x_{3})\Tilde{X}_n(x_{4},x_{4'},x_{2'},x_{3'})}{\omega - (E_n^{N+2} - E_0^N) + i\eta}  \nonumber \\ &
        -i\frac{\Tilde{Z}_n(x_{1},x_2,x_{1'},x_{3})Z_n(x_{4},x_{4'},x_{2'},x_{3'})}{\omega + (E_n^{N-2} - E_0^N) - i\eta}\big],
\end{align}
with 
\begin{align}\label{Eqn:G4_amplitudes}
          X_n(x_{1},x_2,x_{1'},x_{2'})&= \langle\Psi_0 ^N|\hat{\psi}(x_{1})\hat{\psi}(x_2)\hat{\psi}(x_{2'})\hat{\psi}^\dag(x_{1'})|\Psi_n ^{N+2}\rangle , \nonumber \\  
          \Tilde{X}_n(x_{1},x_{2},x_{1'},x_{2'}) &=\langle\Psi_n ^{N+2}|\hat{\psi}(x_{1})\hat{\psi}^\dag(x_2)\hat{\psi}^\dag(x_{2'})\hat{\psi}^\dag(x_{1'})|\Psi_0 ^N\rangle \nonumber\\
          Z_n(x_{1},x_2,x_{1'},x_{2'})&= \langle\Psi_0 ^N|\hat{\psi}(x_{1})\hat{\psi}^\dag(x_2)\hat{\psi}^\dag(x_{2'})\hat{\psi}^\dag(x_{1'})|\Psi_n ^{N-2}\rangle , \nonumber \\  
          \Tilde{Z}_n(x_{1},x_{2},x_{1'},x_{2'}) &=\langle\Psi_n ^{N-2}|\hat{\psi}(x_{1})\hat{\psi}(x_2)\hat{\psi}(x_{2'})\hat{\psi}^\dag(x_{1'})|\Psi_0 ^N\rangle.
\end{align}
From its spectral representation we see that the $G^{3h1e/3e1h}_4$ has the same poles as $G^{2p}_2$ but different amplitudes.  
The $G_2^{2p}$ can be obtained from $G_4^{3h1e/3e1h}$ from the following contraction
\footnote{We note that several choices are possible to integrate out the extra degrees of freedom. This reflects the fact that $G_4$ contains redundant information about $G_2$, as we will discuss later.} 
\begin{align}
    &G_2^{2p}(x_2,x_3,x_{2'},x_{3'},\omega) =-\ \dfrac{1}{(N-1)^2}
    \nonumber \\ &
     \times \int dx_1dx_4G_4^{3h1e/3e1h}(x_1,x_2,x_3,x_4,x_{1},x_{2'},x_{3'},x_{4},\omega),
\end{align}
where we used the fact that
\begin{align}
\int dx_2\hat{\psi}^\dagger(x_2)\hat{\psi}(x_2)|n\rangle=N_n |n\rangle.
\end{align}

In practice it is convenient to project the 4-GF onto a basis.
Using the following transformation 
\begin{align}\label{G4changebasis:eq}
       G_{ijln;mokp}^{3h1e/3e1h}(\omega)&= \int dx_1dx_2dx_3dx_4dx_{1'}dx_{2'}dx_{3'}dx_{4'}\nonumber\\
       &\times \phi_i^*(x_1)\phi_j^*(x_2)\phi^*_l(x_3)\phi_n(x_{1'})\nonumber\\
    &\times G_4^{3h1e/3e1h}(x_1,x_2,x_3,x_4,x_{1'},x_{2'},x_{3'},x_{4'};\omega)\nonumber\\
     &\times\phi^*_p(x_{4})\phi_m(x_{4'})\phi_o(x_{3'})\phi_k(x_{2'}),
\end{align}
the spectral representation in Eq.~\eqref{G4spectral:eq} can be rewritten as
\begin{align}\label{G4basis:eq}
    G_{ijln;mokp}^{3h1e/3e1h}(\omega)&=i\sum_{n'} \frac{X_{n'}^{ijln} \tilde X_{n'}^{mokp}}{\omega-(E_{n'}^{N+2}-E_0^N)+i\eta}\nonumber\\
    &-i\sum_{n'} \frac{\tilde Z_{n'}^{ijln}  Z _{n'}^{mokp}}{\omega+(E_{n'}^{N-2}-E_0^{N})-i\eta},
\end{align}
in which
\begin{align}    \label{G4basis_e:eq}
    X_{n'}^{ijln}&=\langle \Psi_0^N|\hat c_i \hat c_j \hat c_l c^{\dagger}_n |\Psi_{n'}^{N+2}\rangle,\nonumber\\    
    \tilde X_{n'}^{mokp}&=\langle \Psi_{n'}^{N+2}|\hat c_p \hat c^{\dagger}_m \hat c^{\dagger}_o \hat c^{\dagger}_k    |\Psi_0^N\rangle,\nonumber\\
   Z_{n'}^{mokp}&=\langle \Psi_0^N|\hat c_p \hat c^{\dagger}_m \hat c^{\dagger}_o c^{\dagger}_k |\Psi_{n'}^{N-2}\rangle, \nonumber\\
    \tilde Z_{n'}^{ijln}&=\langle \Psi_{n'}^{N-2}|\hat c_i \hat c_j \hat c_l \hat c^{\dagger}_n    |\Psi_0^N\rangle,
\end{align}
where $\hat{c}^{\dagger}$ and $\hat{c}$ are creation and annihilation operators, respectively. In the following we will drop the superscript $3h1e/3e1h$ for notational convenience.
\subsection{Independent-particle $G_4$}
Let us analyse the IP 4-GF since we will need it to solve the MCDE. 
Using Wick's theorem we can write it as the following determinant
\begin{align}\label{G04wick:eq}  
&G^0_{4}(1,2,3,4,1',2',3',4') = 
\nonumber \\
&\begin{vmatrix}
    G^0_1(1,1') & G^0_1(2,1') & G^0_1(3,1') & G^0_1(4,1')\\
    G^0_1(1,2') & G^0_1(2,2') & G^0_1(3,2') & G^0_1(4,2')\\
    G^0_1(1,3') & G^0_1(2,3') & G^0_1(3,3') & G^0_1(4,3')\\
    G^0_1(1,4') & G^0_1(2,4') & G^0_1(3,4') & G^0_1(4,4')\\
    \end{vmatrix},
\end{align}
which generates $4!=24$ terms, each composed of a product of four IP 1-GF. 
Using the time differences defined in Eq.~\eqref{timesG4:eq} we can split the 24 terms in two groups.
In one group each term contains a product $G^0(\tau)G^0(\tau)G^0(\tau)G^0(-\tau)$ and in the other group each term contains a product  $G^0(\tau)G^0(\tau)G^0(0^+)G^0(0^+)$.
We can obtain the spectral representations of these contributions by performing a Fourier transformation with respect to $\tau$.
Using the basis set transformation defined in Eq.~(\ref{G4changebasis:eq}) we obtain two types of spectral representations, one for each group. They are given by
\begin{align}
&[G^0_{i;m}G^0_{j;o}G^0_{l;k}G^0_{p;n}](\omega)
=i\frac{\delta_{im}\delta_{pn}\delta_{jo}\delta_{lk}f^-_{ip}f^-_{jp}f^-_{lp}}{\omega-\Delta\epsilon^+_{ij}-\Delta\epsilon^-_{lp}+i\eta sign(\epsilon_i-\mu)}\label{Eqn:4pcontribution}\\
&- G^0_{i;n}(0^+)G^0_{p;m}(0^+)[G^0_{j;o}G^0_{l;k}](\omega)
\nonumber\\
&=i\frac{\delta_{in}\delta_{pm}\delta_{jo}\delta_{lk}(f^+_{jl}-1)(1-f_i)(1-f_p)}{\omega-\Delta\epsilon^+_{jl}+i\eta sign(\epsilon_j-\mu)}\label{Eqn:2pcontribution},
\end{align}
where $[G^0_{i;m}...G^0_{p;n}](\omega)$ implies a frequency convolution, and $G^0_{i;l}(0^+)=-i(1-f_i)\delta_{il}$. Moreover $\Delta\epsilon^+_{ij}=\epsilon_i+\epsilon_j$, $\Delta\epsilon^-_{lp}=\epsilon_l-\epsilon_p$, $f^-_{ip}=f_i-f_p$, $f^+_{jl}=f_j+f_l$.
We note that the occupation numbers $f^-_{ip}f^-_{jp}f^-_{lp}$ in Eq.~\eqref{Eqn:4pcontribution} restrict this contribution to $G_{4}^0$ to its $3e1h$ and $3h1e$ channels. The indices $i,j,l, m,o, k$ refer to conduction (valence) states and they describe the $3e$ ($3h$) propagation, while $n, p$ refer to valence (conduction) states and they describe the $1h$ ($1e$) propagation. Instead, the occupation numbers $(f^+_{jl}-1)$ in Eq.~\eqref{Eqn:2pcontribution} restrict this contribution to $G_{4}^0$ to its $2e$ and $2h$ channels. 

The pole in the expression on the right-hand side of Eq.~\eqref{Eqn:2pcontribution} is equal to the sum of two energies corresponding to either two valence or two conduction states.
Instead, the pole in the expression on the right-hand side of Eq.~\eqref{Eqn:4pcontribution} is equal to the sum of two energies corresponding to two valence or two conduction states plus the energy difference between a conduction and a valence state.
Therefore, these poles represent, respectively, double electron removal and addition as well as double electron removal and addition accompanied by an electron-hole excitation.
Therefore, the $3h1e$ and $3e1h$ channel of the IP 4-GF contains information about both quasiparticles and satellites.

It can be verified that $G^0_{ijln;mokp}(\omega)$  is invariant upon the permutation of the indices in the triplets ($i,j,l$) and ($m,o,k$) (modulo a minus sign for an odd number of permutations).\footnote{We note that the symmetry in the permutation of the indices is also fulfilled by the interacting $G_4$ as can be seen from Eq.~\eqref{G4basis:eq}.} We can remove this redundant information by restricting the space in which $G_{ijln;mokp}^{0}(\omega)$ is defined with the constraints $i > j > l$ and $m > o > k$.
Moreover, also the following symmetry relation holds

\begin{align}
    \label{G4symmetry:eq}
    -G^{0,3h1e/3e1h}_{cjlc;c'okc'}(\omega) &= G^{0,2p}_{jl;ok}(\omega) \quad (j\neq c, k \neq c' \quad \forall\,c,c')
\end{align}
where $c$ and $c'$ refer to conduction states, i.e., $f_c = f_{c'} = 0$, and $G^{0,2p}_{jl;ok}(\omega) $ is defined according to Eq.~\eqref{Eqn:G2pextactdef}. Finally, $G^{0,2p}_{jl;ok}(\omega)$ itself also contains redundant information, which can be removed using the constraints $j>l$ and $o>k$. \cite{riva_prb} 
With all the restrictions over the space discussed above, and by conveniently defining ${K}_{4}=iG_{4}$, we get 
\begin{align}\label{L04matrix:eq}
  K^0_{4}(\omega) =
    \begin{pmatrix}
    {K}^{\text{0,2p}}(\omega) & 0\\
    0 &  {K}^{\text{0,4p}}(\omega)
    \end{pmatrix},
\end{align}
with

\begin{align}
{K}_{j>l;o>k}^{\text{0,2p}}(\omega)
&=\frac{\delta_{jo}\delta_{lk}(f^+_{jl}-1)}{\Delta\epsilon^+_{jl}-\omega+i\eta sign(\mu-\epsilon_j)}\label{Eqn:K02p},\\
{K}_{i>j>ln;m>o>kp}^{\text{0,4p}}(\omega)&=\frac{\delta_{im}\delta_{pn}\delta_{jo}\delta_{lk}f^-_{ip}f^-_{jp}f^-_{lp}}{\Delta\epsilon^+_{ij}+\Delta\epsilon^-_{lp}-\omega+i\eta sign(\mu-\epsilon_i)}\label{Eqn:K04p}.
\end{align}
It is now clear that the IP 4-GF in the 3e1h/3h1e channel consists of a two-particle channel and a four-particle channel, which are not coupled. 

All the relations we have obtained hold for any 4-GF built from IP 1-GFs. 
In particular, it is convenient to use the Hartree-Fock 1-GF as the IP 1-GF.
Therefore, in the following, it will be understood that $G^0_{1}$ refers to the Hartree-Fock 1-GF.
\subsection{$(4,2)$-MCDE: approximation to the self-energy}\label{multiG2G4:sec}
The IP pp Green's function can be coupled to the 3h1e/3e1h channels of the 4-body Green's function through the following multichannel Dyson equation
\begin{equation}\label{multi_Dyson4:eq}
  K_4(\omega)=K^0_{4}(\omega)+ K^0_{4}(\omega) \Sigma_4  K_4(\omega),
\end{equation}
where $\Sigma_4$ is the multichannel self-energy.
Following Ref.~\cite{riva_prb} we will refer to this MCDE as the $(4,2)$-MCDE, 
where $4$ denotes the highest-order n-body Green's function used and $2$ the change in electron number between the initial and final state.
The multichannel self-energy $\Sigma_4$ is defined by
\begin{equation}\label{self4:eq}
 \Sigma_4=\begin{pmatrix}
        \Sigma^{2\text p} & \Sigma^\text{2p/4p}\\
        \Sigma^\text{4p/2p} & \Sigma^{4\text p}
    \end{pmatrix}.
\end{equation}
where $\Sigma^{2\text p}$ and $\Sigma^{4\text p}$ dress the 2-particle and 4-particle channels, respectively, and $\Sigma^\text{2p/4p}$ and $\Sigma^\text{4p/2p}$ couple the 2-particle and 4-particle channels. 
The approximation that we use for it is analogous to the approximation we have used for the $(3,1)$-MCDE and the $(4,0)$-MCDE.~\cite{riva_prl,riva_prb,riva_prb_25}
We include all contributions in $\Sigma_4$ that are first order in the interaction.
This corresponds to letting each pair of particles interact at the RPAx level, i.e., through a direct and an exchange interaction. 
The head of the matrix in Eq.~\eqref{self4:eq} thus corresponds to the standard RPAx kernel of the BSE, i.e. 
 \begin{equation}\label{Sigma2p}
\Sigma^{\text{2p}}_{jl;ok}=-\overline{\textit{v}}_{jlok}
\end{equation}
where $\bar v_{jlok}=v_{jlok}-v_{jlko}$ with
\begin{equation}
    v_{jlok}=\int dx_1 dx_2 \phi^*_j(x_1)\phi^*_l(x_2)v(\mathbf{r}_1,\mathbf{r}_2)\phi_o(x_2)\phi_k(x_1).\label{potential:eq}
\end{equation}
%
Our approximation yields the following static approximation for $\Sigma^{\text{4p}}$
\begin{align}\label{Sigma4p}
\Sigma^{\text{4p}}_{ijln;mokp}&=- 
\delta_{lk}\delta_{pn}\overline{\textit{v}}_{ijmo}-
\delta_{jo}\delta_{pn}\overline{\textit{v}}_{ilmk}-
\delta_{im}\delta_{pn}\overline{\textit{v}}_{jlok}\nonumber\\
&-
\delta_{lm}\delta_{pn}\overline{\textit{v}}_{ijok}+\delta_{lo}\delta_{pn}\overline{\textit{v}}_{ijmk}+\delta_{jk}\delta_{pn}\overline{\textit{v}}_{ilmo}\nonumber\\
&+\delta_{jm}\delta_{pn}\overline{\textit{v}}_{ilok}-
\delta_{ik}\delta_{pn}\overline{\textit{v}}_{jlmo}+\delta_{io}\delta_{pn}\overline{\textit{v}}_{jlmk}\nonumber\\
&-
\delta_{im}\delta_{jo}\overline{\textit{v}}_{lpnk}+\delta_{io}\delta_{jm}\overline{\textit{v}}_{lpnk}+\delta_{im}\delta_{jk}\overline{\textit{v}}_{lpno}\nonumber\\
&-
\delta_{ik}\delta_{jm}\overline{\textit{v}}_{lpno}-
\delta_{io}\delta_{jk}\overline{\textit{v}}_{lpnm}+\delta_{ik}\delta_{jo}\overline{\textit{v}}_{lpnm}\nonumber\\
&-
\delta_{im}\delta_{lk}\overline{\textit{v}}_{jpno}+\delta_{ik}\delta_{lm}\overline{\textit{v}}_{jpno}+\delta_{im}\delta_{lo}\overline{\textit{v}}_{jpnk}\nonumber\\
&-
\delta_{io}\delta_{lm}\overline{\textit{v}}_{jpnk}+\delta_{io}\delta_{lk}\overline{\textit{v}}_{jpnm}-
\delta_{ik}\delta_{lo}\overline{\textit{v}}_{jpnm}\nonumber\\
&-
\delta_{jo}\delta_{lk}\overline{\textit{v}}_{ipnm}+\delta_{jk}\delta_{lo}\overline{\textit{v}}_{ipnm}+
\delta_{jm}\delta_{lk}\overline{\textit{v}}_{ipno}\nonumber\\
&-
\delta_{jk}\delta_{lm}\overline{\textit{v}}_{ipno}-
\delta_{jm}\delta_{lo}\overline{\textit{v}}_{ipnk}+
\delta_{jo}\delta_{lm}\overline{\textit{v}}_{ipnk}  , 
\end{align} 
while the two coupling terms are given by
\begin{align}\label{Sigma2p/4p}
\Sigma^{\text{2p/4p}}_{jl;mokp}& 
= \overline{\textit{v}}_{jpko}\delta_{lm}
+\overline{\textit{v}}_{jpom}\delta_{lk}
+\overline{\textit{v}}_{jpmk}\delta_{lo}\nonumber\\
&+\overline{\textit{v}}_{lpmo}\delta_{jk}
+\overline{\textit{v}}_{lpok}\delta_{jm}
+\overline{\textit{v}}_{lpkm}\delta_{jo},
\end{align}
and
\begin{align}\label{Sigma4p/2p}
   \Sigma^{\text{4p/2p}}_{ijln;ok}& = 
   \overline{\textit{v}}_{ljon}\delta_{ik}
   +\overline{\textit{v}}_{jion}\delta_{lk}
+\overline{\textit{v}}_{ilon}\delta_{jk}
\nonumber\\
&+\overline{\textit{v}}_{ijkn}\delta_{lo}+
\overline{\textit{v}}_{jlkn}\delta_{io}+
\overline{\textit{v}}_{likn}\delta_{jo}.
\end{align}
We note that $\Sigma_4$ is Hermitian since both $\Sigma^{\text{2p}}$ and  $\Sigma^{\text{4p}}$ are Hermitian and $\Sigma^{\text{2p/4p}} = \left[\Sigma^{\text{4p/2p}}\right]^{\dagger}$.
\subsection{Diagrammatic analysis}
It is useful to represent the $(4,2)$-MCDE in Eq.~\eqref{multi_Dyson4:eq} diagrammatically,  as
\begin{widetext}
\begin{equation}\label{Eqn:Dyson_diag}
    \begin{gathered}
     \includegraphics[width=1\textwidth,clip=]{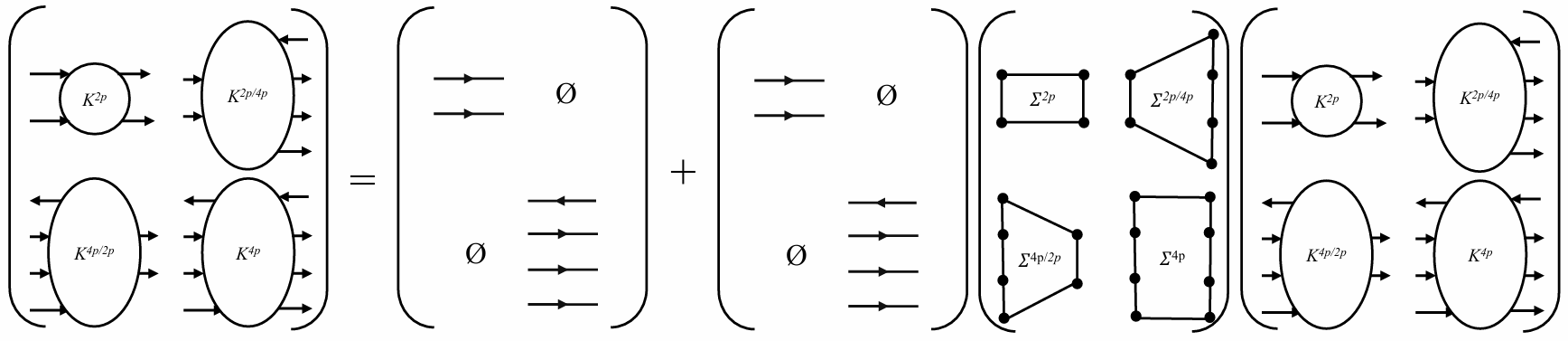}
    \end{gathered}.
\end{equation}
\end{widetext}
Equation ~(\ref{Eqn:Dyson_diag}) shows that $K_4$ consists of the standard two-particle $K^{2\text{p}}$ along with an explicit four-body component $K^{4\text p}$, and the coupling between $K^{2\text{p}}$ and $K^{4\text p}$, denoted as $K^{2\text p/4\text p}$ and $K^{4\text p/2\text p}$. We represent the self-energy coupling terms, $\Sigma^{2\text p/4\text p}_{jl;mokp}$ and $\Sigma^{4\text p/2\text p}_{ijln;ok }$, and the body of the self-energy, $\Sigma^{4\text p}_{ijln;mokp}$,  using isosceles trapezoids and a rectangle, respectively, to reflect their dimensions.

To represent diagrammatically the multichannel self-energy in Eqs.~\eqref{Sigma4p}-\eqref{Sigma4p/2p}, it is convenient to first rewrite them in real space. Using the same change of basis as in Eq.~\eqref{G4changebasis:eq}, these real-space equations are given in Appendix~\ref{sec:realspace}.
The expressions of the $(4,2)$-MCDE in real space makes it easier to understand the diagrammatic structure. 
The head $\Sigma^{2p}$ is the kernel of the $pp$ BSE in the RPAx approximation.\cite{Sch04,Zhang_JPCL2017,Loo22} 
Diagrammatically it can be represented as
\begin{equation}\label{pp-RPAX-kernel}
    \begin{gathered}
        \includegraphics[width=0.47\textwidth,clip=]{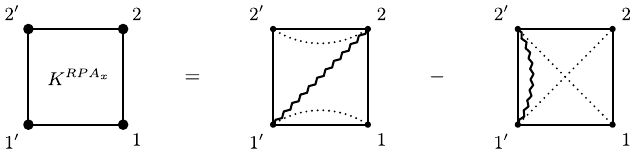},
    \end{gathered}
\end{equation}
where the wiggly lines represent the bare Coulomb interaction and the dotted lines represent a Dirac delta function.
The remaining three components of $\Sigma_4$ are expressed diagrammatically as
\begin{widetext}
\begin{equation}\label{Diagrammes_4p}
\includegraphics[scale = 0.9]{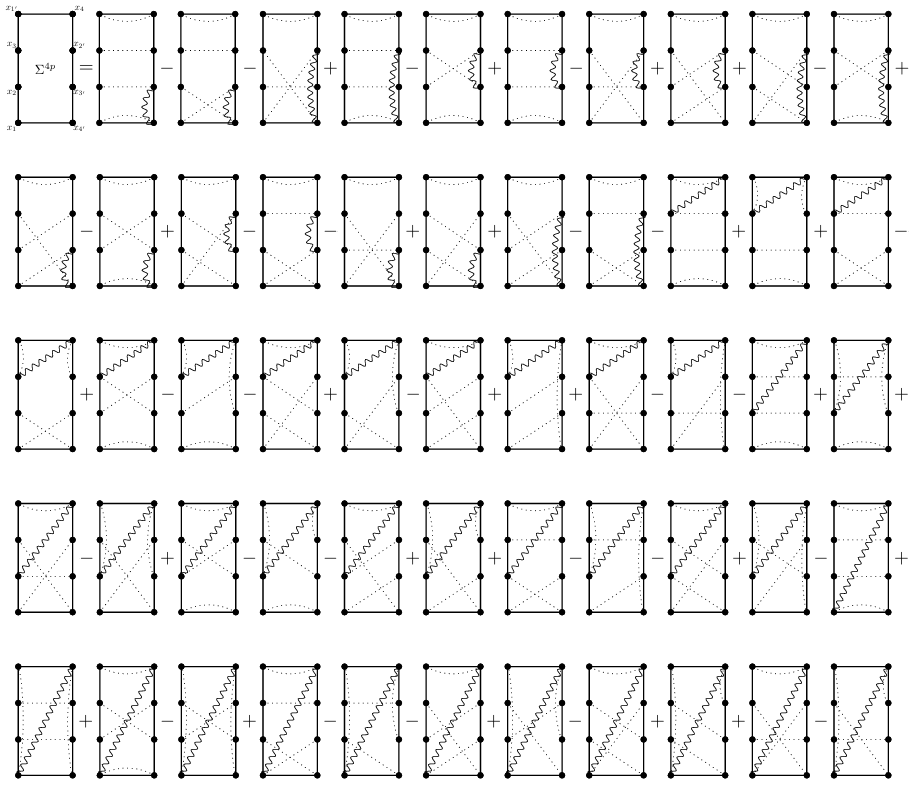}
\end{equation}
\end{widetext}
\begin{widetext}
\begin{equation}\label{Diagrammes_2p-4p}
\includegraphics[scale = 0.9]{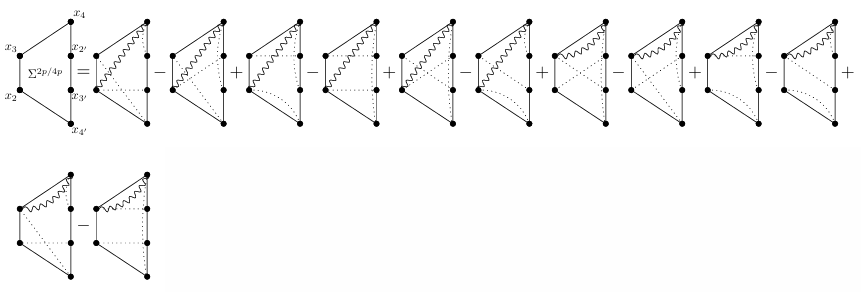}
\end{equation}
\begin{equation}\label{Diagrammes_4p-2p}
\includegraphics[scale = 0.9]{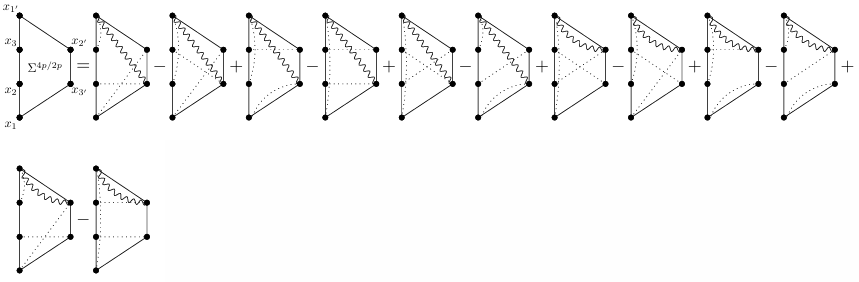}
\end{equation}
\end{widetext}

\begin{figure}[t]
\includegraphics[scale = 0.9]{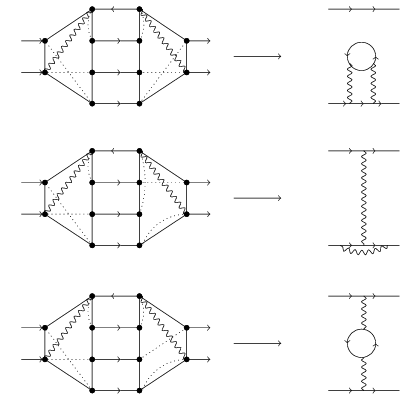}
\caption{Examples of second-order diagrams included in $K^{2p}$ through the approximate $\Sigma_4$.}
\label{Diagrammes_2p}
\end{figure}

To understand the diagrams that are included in $K^{\text{2p}}$ after solving the $(4,2)$-MCDE, we can iterate its diagrammatic expression in Eq.~(\ref{Eqn:Dyson_diag}) and inspect the head of $K_4$ since it corresponds to $K^{\text{2p}}$.
The first iteration is equivalent to the first iteration of the BSE with the RPAx kernel. 
Upon a second iteration, the coupling terms $\Sigma^{\text{2p/4p}}$ and  $\Sigma^{\text{4p/2p}}$ will contribute to the head of $K_4$. 
They add correlation in three different ways: 1) they dress single $G^0_{1}$ lines (see upper diagram in Fig.~\ref{Diagrammes_2p} for an example). As a consequence, the IP particle-particle Green's function in Eq.~\eqref{Eqn:Dyson_diag} should not be evaluated beyond the Hartree-Fock approximation as this would lead to a double counting of diagrams and, therefore, of correlation;
2) they create mixed terms in which a dressed $G^0_{1}$ interacts with a bare $G^0_{1}$ (see middle diagram in  Fig.~\ref{Diagrammes_2p} for an example); 3) they dress the interaction between the two particles, for example by screening the interaction (see lower diagram in  Fig.~\ref{Diagrammes_2p} for an example). 
We note that after two iterations of the $(4,2)$-MCDE with the coupling terms in Eqs.~\eqref{Diagrammes_2p-4p} and~\eqref{Diagrammes_4p-2p}, we obtain all diagrams that are second-order in the interaction.
In other words, the $(4,2)$-MCDE is exact up to second order in the interaction. 

\begin{figure}[t]
\includegraphics[scale = 0.3]{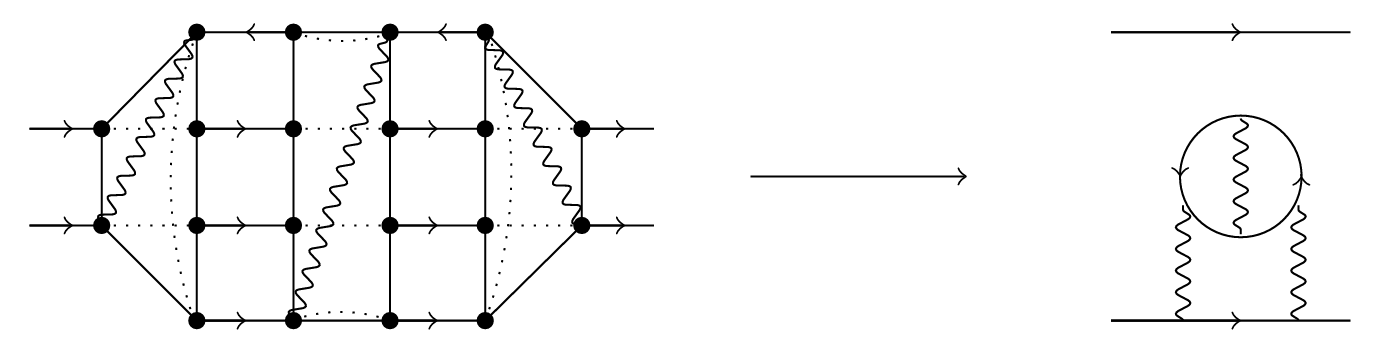}
\caption{Example of third-order diagram included in $K^{2p}$ through the approximate $\Sigma_4$.}
\label{3rd order diagram}
\end{figure}

After a third iteration of the $(4,2)$-MCDE, also the body of $\Sigma_4$ contributes to the diagrams (see Fig.(\ref{3rd order diagram}) for an example). 
Similarly to the $(3,1)$- and $(4,0)$-MCDE, also in the case of the $(4,2)$-MCDE one can further dress the final $K^{\text{2p}}$ by using a $K^{\text{0,4p}}$ beyond HF.  To go beyond the RPAx approximation for $\Sigma_4$, one may include screening effects in the equations. To avoid double counting, screening should be incorporated only in $\Sigma^{4p}$ and in all terms except the electron-hole exchange interactions. We have already included static screening in the multichannel Dyson equation describing the coupling between the one- and three-body Green's functions, resulting in a noticeable improvement in the description of the plasmon satellite in bulk silicon. \cite{berger2026}

Finally, it is instructive to express the particle-particle correlation self-energy $\Sigma^{\text{2p,c}}(\omega)$ in terms of $\Sigma_4$. It reads 
\begin{equation}
\Sigma^{\text{2p,c}}_{jl;ok}(\omega)=\Sigma^{\text{2p/4p}}_{jl;mnrs}K^{\text{4p}}_{mnrs;m'n'r's'}(\omega)\Sigma^{\text{4p/2p}}_{m'n'r's';ok}
\label{Eqn:downfold}
\end{equation}
where
\begin{widetext}
 \begin{equation}
K^{\text{4p}}_{mnrs;m'n'r's'}(\omega)=[K^{\text{4p},-1}_{\bar{m}\bar{n}\bar{r}\bar{s};\bar{m'}\bar{n'}\bar{r'}\bar{s'}}(\omega)\nonumber\\-\Sigma^{4\text p}_{\bar{m}\bar{n}\bar{r}\bar{s};\bar{m'}\bar{n'}\bar{r'}\bar{s'}}]^{-1}_{mnrs;m'n'r's'}
\label{Eqn:K4}.
\end{equation}
\end{widetext}
The details of the derivation are given in App.~\ref{CorrelatedS}.
The above expressions demonstrate that:
1) even though $\Sigma_4$ is static, the corresponding 2-body pp self-energy is dynamical thanks to $K^{\text{4p}}(\omega)$;
2) since $\Sigma^{\text{2p/4p}} = [\Sigma^{\text{4p/2p}}]^{\dagger}$, the $(4,2)$-MCDE is guaranteed to yield a positive-definite spectrum~\cite{Ste14};
3) although the multichannel self-energy $\Sigma_4$ only contains contributions that are of first order in the interaction, thanks to the inverse matrix in Eq.~\ref{Eqn:K4}, 
the corresponding 2-body particle-particle self-energy contains many contributions that are of infinite order in the interaction.
Finally, we note that approximations to $\Sigma^{\text{2p,c}}$ can be rewritten in the form given in Eq.~\ref{Eqn:downfold}~\cite{Marie_2025} .

%

\subsection{Effective four-particle Hamiltonian}
In order to solve the Dyson equation (\ref{multi_Dyson4:eq}) in practice, it is convenient to express it as an effective 4-particle Hamiltonian by using
\begin{equation}
K_4(\omega)=\left[[K^{0}_{4}(\omega)]^{-1}-\Sigma_4\right]^{-1}.
\end{equation}

Therefore, $K_4(\omega)$ can be written as
\begin{widetext}
\begin{align}\label{L-1}
 K_4(\omega)&=\begin{pmatrix}
       \frac{\delta_{jo}\delta_{lk}\left(\Delta^+\epsilon_{jl} -\omega\right)}{(f^+_{jl}-1)}-\Sigma^{2\text p}_{j>l;o>k} & -\Sigma^\text{2p/4p}_{j>l;m>o>kp}\\
       - \Sigma^\text{4p/2p}_{i>j>ln;o>k} &  \frac{\delta_{im}\delta_{jo}\delta_{lk}\delta_{np} \left(\Delta^-\epsilon_{in} + \Delta^+\epsilon_{jl}-\omega\right)}{f^-_{ip}f^-_{jp}f^-_{lp}}{}-\Sigma^{4\text p}_ {i>j>ln;m>o>kp}
    \end{pmatrix}^{-1}.
 \end{align}
%
\\
Using $[AB]^{-1} = B^{-1}A^{-1}$ we obtain

\begin{equation}\label{Eqn:L4_0}
K_4(\omega)=\left[H^{\text{eff}}_4(\omega)-\omega\mathbf{1}\right]^{-1}\begin{pmatrix}
   \delta_{jo}\delta_{lk}(f^+_{jl}-1)&0\\
    0&  \delta_{im}\delta_{jo}\delta_{lk}\delta_{np}f^-_{in}f^-_{jn}f^-_{ln} \end{pmatrix},
 \end{equation}
where the effective Hamiltonian is defined as
\begin{align}\label{Eqn:H4}
H^{\text{eff}}_4=\begin{pmatrix}
     \delta_{jo}\delta_{lk}\Delta^+\epsilon_{jl}-(f^+_{jl}-1)\Sigma^{2\text p}_{j>l;o>k}& -(f^+_{jl}-1)\Sigma^\text{2p/4p}_{j>l;m>o>k p}\\
       -f^-_{in}f^-_{jn}f^-_{ln}\Sigma^\text{4p/2p}_{i>j>ln;o>k} & \delta_{im}\delta_{jo}\delta_{lk}\delta_{np} \left(\Delta^-\epsilon_{in} + \Delta^+\epsilon_{jl}\right)-f^-_{in}f^-_{jn}f^-_{ln}\Sigma^{4\text p}_ {i>j>ln;m>o>kp}             
       \end{pmatrix}.
\end{align}
\end{widetext}
%


The presence of the factors $(f^+_{jl}-1)$ and $f^-_{in}f^-_{jn}f^-_{ln}$ in Eq.~\eqref{Eqn:L4_0} ensures that 
only the particle-particle and the 3-hole-1-electron and 3-electron-1-hole channels are included in the Hamiltonian.
These factors also account for the correct sign of each contribution. 
Since the effective Hamiltonian is static and Hermitian, we can now write

\begin{equation}\label{Eqn:eigenvaluep}
    H^{\text{eff}}_4A_\lambda = E_\lambda A_\lambda,
\end{equation}
where $E_\lambda$ and $A_\lambda$ are the eigenvalues and the eigenvectors, respectively, of $H^{eff}_4$. 
Thus, the spectral representation of the effective Hamiltonian can be written as
\begin{equation}
    [H^{\text{eff}}_4-\omega\mathbf{1}]^{-1}_{\mu\nu}=\sum_{\lambda\lambda'}\frac{A_\lambda^\mu S^{-1}_{\lambda\lambda'}A_{\lambda'}^{*\nu}}{E_\lambda-\omega},
\end{equation}
where $S_{\lambda\lambda'}=\sum_\mu A^{*\mu}_\lambda A^{\mu}_{\lambda'}$ is the
the overlap matrix. 
Therefore, we obtain the following matrix form for $K_4(\omega)$
\begin{equation}
    K_4(\omega) =
    \begin{pmatrix}
        K^{2\text p}(\omega) & K^{\text{2p/4p}}(\omega) \\
        K^{\text{4p/2p}}(\omega) & K^{\text{4p}}(\omega)
    \end{pmatrix},
\end{equation}
where the components are given by
\begin{equation}\label{Eqn:K2p}
    K_{j>l;o>k}^{2\text p}(\omega) = \sum_{\lambda\lambda'}\frac{A^{jl}_\lambda S^{-1}_{\lambda\lambda'} A^{*ok}_{\lambda'}}{E_\lambda - \omega} (f^+_{ok}-1)
\end{equation}
\begin{equation}
    K_{j>l;m>o>kp}^{\text{2p/4p}}(\omega) = \sum_{\lambda\lambda'}\frac{A^{jl}_\lambda S^{-1}_{\lambda\lambda'} A^{*mokp}_{\lambda'}}{E_\lambda - \omega} f_{mp}f_{op}f_{kp}
\end{equation}
\begin{equation}
    K_{i>j>ln;o>k}^{\text{4p/2p}}(\omega) = \sum_{\lambda\lambda'}\frac{A^{ijln}_\lambda S^{-1}_{\lambda\lambda'}A^{*ok}_{\lambda'}}{E_\lambda - \omega} (f^+_{ok}-1)
\end{equation}
\begin{equation}
    K_{i>j>ln;m>o>kp}^{4\text p}(\omega) = \sum_{\lambda\lambda'}\frac{A^{ijln}_\lambda S^{-1}_{\lambda\lambda'}A^{*mokp}_{\lambda'}}{E_\lambda - \omega} f_{mp}f_{op}f_{kp}.
\end{equation}

We can thus obtain these four contributions by solving the eigenvalue problem in Eq.\eqref{Eqn:H4}. 
The spectrum of two-electron removal and two-electron addition energies is obtained from Eq.\eqref{Eqn:K2p}, which is hence the equation of interest here. 
The eigenvalue problem in Eq.\eqref{Eqn:eigenvaluep} can be solved by direct diagonalization or by more efficient iterative techniques.
In particular, using the Haydock-Lanczos method we can directly
solve for the spectral function~\cite{Hay72,Schm03,Her05}. The numerical scaling of the calculation will then be determined by the construction of $H^{\text{eff}}_4$ which scales as $N^6$ with $N$ the number of electrons.

\section{Conclusions and Perspectives \label{Conclusions}}
We derived a multichannel Dyson equation to simulate photoemission spectroscopies involving double ionization and double addition.
Moreover, we derived a simple approximation to the corresponding multichannel self-energy that can describe both quasiparticles and satellites.
Our approximation, therefore, goes well beyond the RPA since it combines several correlation effects that are usually treated separately when using GW- and GT-based kernels. 
We also derived a practical protocol that maps the mutlichannel Dyson equation onto an eigenvalue problem with an effective Hamiltonian.
The resulting eigenvalues and eigenvectors can be used to construct the spectral representation of the particle-particle Green's function.
The MCDE is a general concept, which has already been used to describe photoemission and absorption spectra with very promising results~\cite{riva_prl,paggi_2025}.
We have recently implemented the (3,1)-MCDE to simulate photoemission spectra in real systems \cite{berger_2026} and implementation of the (4,0)- and (4,2)-MCDE is currently in progress.

\textit{Acknowledgment:}
We thank the French "Agence Nationale de la Recherche (ANR)" for financial support (Grant Agreements No. ANR-22-CE30-0027 and No. ANR-
22-CE29-0001).
\appendix
\begin{widetext}

\section{$\Sigma_4$ in the real space\label{sec:realspace}}
In the following we give the expressions of $\Sigma_4$ in real space.
\begin{align}
    \Sigma^{4p}(x_1,x_2,x_3,x_4,x_{1'},x_{2'},x_{3'},x_{4'})&= -\delta(x_{3},x_{2'})\delta(x_{4},x_{1'})[\delta(x_{1},x_{3'})\delta(x_{2},x_{4'})-\delta(x_{1},x_{4'})\delta(x_{2},x_{3'})]v(\textbf{r}_{4'},\textbf{r}_{3'})\nonumber\\
    &-\delta(x_{2},x_{3'})\delta(x_{4},x_{1'})[\delta(x_{1},x_{2'})\delta(x_{3},x_{4'})-\delta(x_{1},x_{4'})\delta(x_{3},x_{2'})]v(\textbf{r}_{4'},\textbf{r}_{2'})\nonumber\\
    &-\delta(x_{3},x_{4'})\delta(x_{4},x_{1'})[\delta(x_{1},x_{2'})\delta(x_{2},x_{3'})-\delta(x_{1},x_{3'})\delta(x_{2},x_{2'})]v(\textbf{r}_{3'},\textbf{r}_{2'})\nonumber\\
    &+\delta(x_{3},x_{4'})\delta(x_{4},x_{1'})[\delta(x_{1},x_{2'})\delta(x_{2},x_{4'})-\delta(x_{1},x_{4'})\delta(x_{2},x_{2'})]v(\textbf{r}_{4'},\textbf{r}_{2'})\nonumber\\
    &+\delta(x_{2},x_{2'})\delta(x_{4},x_{1'})[\delta(x_{1},x_{3'})\delta(x_{3},x_{4'})-\delta(x_{1},x_{4'})\delta(x_{3},x_{3'})]v(\textbf{r}_{4'},\textbf{r}_{3'})\nonumber\\
    &+\delta(x_{2},x_{4'})\delta(x_{4},x_{1'})[\delta(x_{1},x_{2'})\delta(x_{3},x_{3'})-\delta(x_{1},x_{3'})\delta(x_{3},x_{2'})]v(\textbf{r}_{3'},\textbf{r}_{2'})\nonumber\\
    &-\delta(x_{1},x_{2'})\delta(x_{4},x_{1'})[\delta(x_{2},x_{3'})\delta(x_{3},x_{4'})-\delta(x_{2},x_{4'})\delta(x_{3},x_{3'})]v(\textbf{r}_{4'},\textbf{r}_{3'})\nonumber\\
    &+\delta(x_{1},x_{3'})\delta(x_{4},x_{1'})[\delta(x_{2},x_{2'})\delta(x_{3},x_{4'})-\delta(x_{2},x_{4'})\delta(x_{3},x_{2'})]v(\textbf{r}_{4'},\textbf{r}_{2'})\nonumber\\
    &-\delta(x_{1},x_{4'})\delta(x_{2},x_{3'})[\delta(x_{3},x_{2'})\delta(x_{4},x_{1'})-\delta(x_{3},x_{1'})\delta(x_{4},x_{2'})]v(\textbf{r}_{3},\textbf{r}_{4})\nonumber\\
    &+\delta(x_{1},x_{3'})\delta(x_{2},x_{4'})[\delta(x_{3},x_{2'})\delta(x_{4},x_{1'})-\delta(x_{3},x_{1'})\delta(x_{4},x_{2'})]v(\textbf{r}_{3},\textbf{r}_{4})\nonumber\\
    &+\delta(x_{1},x_{4'})\delta(x_{2},x_{2'})[\delta(x_{3},x_{3'})\delta(x_{4},x_{1'})-\delta(x_{3},x_{1'})\delta(x_{4},x_{3'})]v(\textbf{r}_{3},\textbf{r}_{4})\nonumber\\
    &-\delta(x_{1},x_{2'})\delta(x_{2},x_{4'})[\delta(x_{3},x_{3'})\delta(x_{4},x_{1'})-\delta(x_{3},x_{1'})\delta(x_{4},x_{3'})]v(\textbf{r}_{3},\textbf{r}_{4})\nonumber\\
    &-\delta(x_{1},x_{3'})\delta(x_{2},x_{2'})[\delta(x_{3},x_{4'})\delta(x_{4},x_{1'})-\delta(x_{3},x_{1'})\delta(x_{4},x_{4'})]v(\textbf{r}_{3},\textbf{r}_{4})\nonumber\\
    &+\delta(x_{1},x_{2'})\delta(x_{2},x_{3'})[\delta(x_{3},x_{4'})\delta(x_{4},x_{1'})-\delta(x_{3},x_{1'})\delta(x_{4},x_{4'})]v(\textbf{r}_{3},\textbf{r}_{4})\nonumber\\
    &-\delta(x_{1},x_{4'})\delta(x_{3},x_{2'})[\delta(x_{2},x_{3'})\delta(x_{4},x_{1'})-\delta(x_{2},x_{1'})\delta(x_{4},x_{3'})]v(\textbf{r}_{2},\textbf{r}_{4})\nonumber\\
    &+\delta(x_{1},x_{2'})\delta(x_{3},x_{4'})[\delta(x_{2},x_{3'})\delta(x_{4},x_{1'})-\delta(x_{2},x_{1'})\delta(x_{4},x_{3'})]v(\textbf{r}_{2},\textbf{r}_{4})\nonumber\\
    &+\delta(x_{1},x_{4'})\delta(x_{3},x_{3'})[\delta(x_{2},x_{2'})\delta(x_{4},x_{1'})-\delta(x_{2},x_{1'})\delta(x_{4},x_{2'})]v(\textbf{r}_{2},\textbf{r}_{4})\nonumber\\
    &-\delta(x_{1},x_{3'})\delta(x_{3},x_{4'})[\delta(x_{2},x_{2'})\delta(x_{4},x_{1'})-\delta(x_{2},x_{1'})\delta(x_{4},x_{2'})]v(\textbf{r}_{2},\textbf{r}_{4})\nonumber\\
    &+\delta(x_{1},x_{3'})\delta(x_{3},x_{2'})[\delta(x_{2},x_{4'})\delta(x_{4},x_{1'})-\delta(x_{2},x_{1'})\delta(x_{4},x_{4'})]v(\textbf{r}_{2},\textbf{r}_{4})\nonumber\\
    &-\delta(x_{1},x_{2'})\delta(x_{3},x_{3'})[\delta(x_{2},x_{4'})\delta(x_{4},x_{1'})-\delta(x_{2},x_{1'})\delta(x_{4},x_{4'})]v(\textbf{r}_{2},\textbf{r}_{4})\nonumber\\
    &-\delta(x_{2},x_{3'})\delta(x_{3},x_{2'})[\delta(x_{1},x_{4'})\delta(x_{4},x_{1'})-\delta(x_{1},x_{1'})\delta(x_{4},x_{4'})]v(\textbf{r}_{1},\textbf{r}_{4})\nonumber\\
    &+\delta(x_{2},x_{2'})\delta(x_{3},x_{3'})[\delta(x_{1},x_{4'})\delta(x_{4},x_{1'})-\delta(x_{1},x_{1'})\delta(x_{4},x_{4'})]v(\textbf{r}_{1},\textbf{r}_{4})\nonumber\\
    &+\delta(x_{2},x_{4'})\delta(x_{3},x_{2'})[\delta(x_{1},x_{3'})\delta(x_{4},x_{1'})-\delta(x_{1},x_{1'})\delta(x_{4},x_{3'})]v(\textbf{r}_{1},\textbf{r}_{4})\nonumber\\
    &-\delta(x_{2},x_{2'})\delta(x_{3},x_{4'})[\delta(x_{1},x_{3'})\delta(x_{4},x_{1'})-\delta(x_{1},x_{1'})\delta(x_{4},x_{3'})]v(\textbf{r}_{1},\textbf{r}_{4})\nonumber\\
    &-\delta(x_{2},x_{4'})\delta(x_{3},x_{3'})[\delta(x_{1},x_{2'})\delta(x_{4},x_{1'})-\delta(x_{1},x_{1'})\delta(x_{4},x_{2'})]v(\textbf{r}_{1},\textbf{r}_{4})\nonumber\\
    &+\delta(x_{2},x_{3'})\delta(x_{3},x_{4'})[\delta(x_{1},x_{2'})\delta(x_{4},x_{1'})-\delta(x_{1},x_{1'})\delta(x_{4},x_{2'})]v(\textbf{r}_{1},\textbf{r}_{4})
\end{align}

\begin{align}
    \Sigma^{2p/4p}(x_2,x_3,x_4,x_{2'},x_{3'},x_{4'})&= \delta(x_{3},x_{3'})[\delta(x_{2},x_{2'})\delta(x_{4},x_{4'})-\delta(x_{2},x_{4'})\delta(x_{4},x_{2'})]v(\textbf{r}_4,\textbf{r}_2)\nonumber\\
    &-\delta(x_{3},x_{4'})[\delta(x_{2},x_{2'})\delta(x_{4},x_{3'})-\delta(x_{2},x_{3'})\delta(x_{4},x_{2'})]v(\textbf{r}_4,\textbf{r}_2)\nonumber\\
    &-\delta(x_{3},x_{2'})[\delta(x_{2},x_{3'})\delta(x_{4},x_{4'})-\delta(x_{2},x_{4'})\delta(x_{4},x_{3'})]v(\textbf{r}_4,\textbf{r}_2)\nonumber\\
    &-\delta(x_{2},x_{3'})[\delta(x_{3},x_{2'})\delta(x_{4},x_{4'})-\delta(x_{3},x_{4'})\delta(x_{4},x_{2'})]v(\textbf{r}_4,\textbf{r}_3)\nonumber\\
    &+\delta(x_{2},x_{2'})[\delta(x_{3},x_{3'})\delta(x_{4},x_{4'})-\delta(x_{3},x_{4'})\delta(x_{4},x_{3'})]v(\textbf{r}_4,\textbf{r}_3)\nonumber\\
    &+\delta(x_{2},x_{4'})[\delta(x_{3},x_{2'})\delta(x_{4},x_{3'})-\delta(x_{3},x_{2'})\delta(x_{4},x_{3'})]v(\textbf{r}_4,\textbf{r}_3)    
\end{align}
\begin{align}
    \Sigma^{4p/2p}(x_1,x_2,x_3,x_{1'},x_{2'},x_{3'})&= \delta(x_{1},x_{2'})[\delta(x_{3},x_{1'})\delta(x_{2},x_{3'})-\delta(x_{3},x_{3'})\delta(x_{2},x_{1'})]v(\textbf{r}_{1'},\textbf{r}_{3'})\nonumber\\
    &+\delta(x_{3},x_{2'})[\delta(x_{2},x_{1'})\delta(x_{1},x_{3'})-\delta(x_{2},x_{3'})\delta(x_{1},x_{1'})]v(\textbf{r}_{1'},\textbf{r}_{3'})\nonumber\\
    &+\delta(x_{2},x_{2'})[\delta(x_{3},x_{3'})\delta(x_{1},x_{1'})-\delta(x_{3},x_{1'})\delta(x_{1},x_{3'})]v(\textbf{r}_{1'},\textbf{r}_{3'})\nonumber\\
    &+\delta(x_{3},x_{3'})[\delta(x_{2},x_{2'})\delta(x_{1},x_{1'})-\delta(x_{2},x_{1'})\delta(x_{1},x_{2'})]v(\textbf{r}_{1'},\textbf{r}_{2'})\nonumber\\
    &+\delta(x_{1},x_{3'})[\delta(x_{3},x_{2'})\delta(x_{2},x_{1'})-\delta(x_{3},x_{1'})\delta(x_{2},x_{3'})]v(\textbf{r}_{1'},\textbf{r}_{2'})\nonumber\\
    &+\delta(x_{2},x_{3'})[\delta(x_{1},x_{2'})\delta(x_{3},x_{1'})-\delta(x_{1},x_{1'})\delta(x_{3},x_{2'})]v(\textbf{r}_{1'},\textbf{r}_{2'})    
\end{align}
\section{Frequency-dependent two-body self-energy from $\Sigma_4$\label{CorrelatedS}}
The procedure described in the following is very general and it can be applied to any Multichannel Dyson Equation to get the corresponding lower-space frequency-dependent self-energy. The starting point is the eigenvalue equation (\ref{Eqn:eigenvaluep}), which can be written schematically as 
 \begin{equation}
	\left(
	\begin{array}{cc}
		H^{2\text p} & {H}^{\text{2p/4p}} \\
	H^{\text{4p/2p}} & H^{4\text p}
	\end{array}
	\right)\left(
	\begin{array}{c}
		A_\lambda^{2\text p} \\
	 A_\lambda^{4\text p}
	\end{array}
	\right)=E_\lambda\left(
	\begin{array}{c}
		A_\lambda^{2\text p} \\
	 A_\lambda^{4\text p}
	\end{array}
	\right)
\end{equation}
where 
\begin{align}
\label{Eqn:H1p}
    H^{2\text p}_{j>l;o>k}&=   \delta_{jo}\delta_{lk}\Delta^+\epsilon_{jl}-(f^+_{jl}-1)\Sigma^{2p}_{j>l;o>k}
    \\
    H^{\text{2p/4p}}_{j>l;m>o>kp}&= -(f^+_{jl}-1)\Sigma^\text{2p/4p}_{j>l;m>o>k p},
    \\
   H^{\text{4p/2p}}_{i>j>ln;ok}& =  -f^-_{in}f^-_{jn}f^-_{ln}\Sigma^\text{4p/2p}_{i>j>ln;o>k},
    \\
    H^{4\text p}_{i>j>ln;m>o>kp}&= \delta_{im}\delta_{jo}\delta_{lk}\delta_{np} \left(\Delta^-\epsilon_{in} + \Delta^+\epsilon_{jl}\right)-f^-_{in}f^-_{jn}f^-_{ln}\Sigma^{4\text p}_ {i>j>ln;m>o>kp}.
    \label{Eqn:H3p}
\end{align}

Let us downfold the problem in the $2p$ space as follows
\begin{align}k
H^{2\text p} A_\lambda^{2\text p}+H^{\text{2p/4p}} A_\lambda^{4\text p}&=E_\lambda A_\lambda^{2\text p}\\
H^{\text{4p/2p}} A_\lambda^{2\text p}+H^{4\text p} A_\lambda^{4\text p}&=E_\lambda A_\lambda^{4\text p}  \Rightarrow  A_\lambda^{4\text p}=\left[E_\lambda-H^{4\text p} \right]^{-1}H^{\text{4p/2p}} A_\lambda^{2\text p}
\end{align}
from which
\begin{align}
H^{2\text p} A_\lambda^{2\text p}+H^{\text{2p/4p}} \underbrace{\left[E_\lambda-H^{4\text p} \right]^{-1}}_{M^{-1}}H^{\text{4p/2p}} A_\lambda^{2\text p}
=E_\lambda A_\lambda^{2\text p}\\
\end{align}

Therefore the 2p effective hamiltonian reads
\begin{align}
\tilde{H}^{\text{eff}}_{jl;ok} &=H_{jl;ok}+H_{jl;mnrs} M^{-1}_{mnrs;m'n'r's'}H_{m'n'r's';ok}\nonumber\\
&=\delta_{jo}\delta_{lk}\Delta^+\epsilon_{jl}-(f^+_{jl}-1)\Sigma^{2p}_{j>l;o>k} -(f^+_{jl}-1)\Sigma^\text{2p/4p}_{j>l;m>n>r s}\nonumber\\
&\times[\delta_{\bar{m}\bar{m}'}\delta_{\bar{n}\bar{n}'}\delta_{\bar{r}\bar{r}'}\delta_{\bar{s}\bar{s}'} \left(\omega-\Delta^-\epsilon_{\bar{m}\bar{s}} - \Delta^+\epsilon_{\bar{n}\bar{r}}\right)+f^-_{\bar{m}\bar{s}}f^-_{\bar{n}\bar{s}}f^-_{\bar{r}\bar{s}}\Sigma^{4\text p}_ {\bar{m}>\bar{n}>\bar{r}\bar{s};\bar{m}'>\bar{n}'>\bar{r}'\bar{s}'} ]^{-1}_{mnrs;m'n'r's'}f^-_{m's'}f^-_{n's'}f^-_{s'r'}\Sigma^\text{4p/2p}_{m'>n'>r's';o>k}\nonumber\\
&=\delta_{jo}\delta_{lk}\Delta^+\epsilon_{jl}-(f^+_{jl}-1)\left[\Sigma^{2p}_{j>l;o>k}+\Sigma^\text{2p/4p}_{j>l;m>n>r s}K^{\text{4p}}_{m>n>rs;m'>n'>r's'}(\omega)\Sigma^\text{4p/2p}_{m'>n'>r's';o>k}\right]
\end{align}

Since
\begin{align}
K^{\text{2p},-1}_{j>l;o>k}(\omega)=\frac{(\Delta\epsilon^+_{jl}-\omega)\delta_{jo}\delta_{lk}}{(f^+_{jl}-1)}-\Sigma_{j>l;o>k}(\omega)
\end{align}
we can identify 
\begin{align}
\Sigma_{j>l;o>k}(\omega)=\Sigma^{\text{2p}}_{j>l;o>k}+\Sigma^{\text{c}}_{jl;ok}(\omega)= \Sigma^{2p}_{j>l;o>k}+\Sigma^\text{2p/4p}_{j>l;m>n>r s}K^{\text{4p}}_{m>n>r s;m'>n'>r's'}(\omega)\Sigma^\text{4p/2p}_{m'>n'>r's';o>k}
\end{align}

\end{widetext} 
%
\end{document}